\documentclass[11pt]{article}
\usepackage{amssymb,amsmath,graphicx}

\newtheorem{theorem}{Theorem}[section]
\newtheorem{proposition}[theorem]{Proposition}
\newtheorem{lemma}[theorem]{Lemma}
\newtheorem{corollary}[theorem]{Corollary}

\newtheorem{remark}[theorem]{Remark}

\newcommand{\eproof}{\rule{0,2cm}{0,2cm}}

\begin{document}

\title{{\Large {\textbf{Functional-differential equations for the $q$%
-Fourier transform of $q$-Gaussians}}}}
\author{{Sabir Umarov} {$^{1}$} and {S\'{\i}lvio M. Duarte Queir\'{o}s} {$%
^{2}$}}
\date{}
\maketitle

\begin{center}
$^{1}$ \textit{Department of Mathematics Tufts University, Medford, MA, USA}%
\\[0pt]
$^{2}$ \textit{Unilever R\&D Port Sunlight, Wirral, UK}\\[0pt]
\end{center}

\begin{abstract}
In the paper the question - \textit{Is the $q$-Fourier transform of a $q$-Gaussian a $%
q^{^{\prime }}$-Gaussian (with some $q^{^{\prime }}$) up to a
constant factor?} - is studied for the whole range of $q\in (-\infty
,3).$ This question is connected with applicability of the
${q}$-Fourier transform in the study of limit processes in
nonextensive statistical mechanics.
We prove that the answer is affirmative
if and only if $q\geq 1,$ excluding two particular cases of $q<1,$
namely, $q=\frac{1}{2}$ and $q=\frac{2}{3},$ which are also out of
the theory valid for $q \ge 1.$ We also discuss some applications of
the $q$-Fourier transform to nonlinear partial differential
equations such as the porous medium equation.
\end{abstract}

\section{Introduction}

Approximately one century after Boltzmann's seminal works which have
turn into the cornerstones of statistical mechanics,
Tsallis~\cite{Tsallis88} introduced an entropic form aimed to
accommodate the description of systems whose fundamental features do
not fit for the properties assumed in the Boltzmann-Bibbs formalism;
~ see \cite{Tsallis09,Tsallis1,Tsallis2}. Tsallis' entropic form,
which is usually called non-additive $q$-entropy, recovers the
classic Boltzmann-Gibbs entropic form
in limit the case $q\rightarrow 1$. Concomitantly, there is the
nonextensive statistical mechanics formalism based on the
$q$-algebra and the $q$-Gaussian probability density function, which
maximises $q$-entropy under certain appropriate constraints (see
\cite{Tsallis2,Tsallis3,Tsallis2005} and
references therein). Recently, the $q$-Fourier transform was
introduced \cite{UmarovTsallisSteinberg} as a tool for the study of
attractors of strongly correlated random variables arising in
nonextensive statistical mechanics.
In this paper, we shed light on the question - whether the
${q}$-Fourier transform of a $q$-Gaussian is a $q^{^{\prime
}}$-Gaussian for some another $q^{\prime }$ again. A key to this
question is crucial because, as a mathematical tool, the $F_{q}$ is
relevant to both the study of limit distributions and, as we will show
later on this paper, the solution of partial differential equations
with physical significance as well. Moreover, a positive answer
implies validating the mapping relation of $q$ onto $q^{\prime }$ obtained from $F_{q}$,
which has been predominant for the establishment of other stable
distributions, namely the $\left( q, \alpha \right)$-stable
distributions \cite{UmarovTsallisGell-MannSteinberg}.
We recall that, by definition, the \textit{$F_q$-transform}, or
\textit{$q$-Fourier transform} of a nonnegative $f\in L_1(R) $ is
defined by the formula
\begin{equation}  \label{Fourier}
F_q[f](\xi) = \int_{supp \, f} e_q^{ix\xi} \otimes_q f(x) dx \, ,
\end{equation}
where $q<3,$ the symbol $\otimes_q$ stands for the $q$-product, and
\begin{equation}
e_q^{z}=(1+(1-q)z)^{1/(1-q)}, \, z\in C,
\end{equation}
is a $q$-exponential (see \cite{Tsallis2005,UmarovTsallisSteinberg} for details). The equality
\begin{equation*}
e_q^{ix\xi} \otimes_q f(x) = f(x) e_q^{\frac{ix \xi}{[f(x)]^{1-q}}},
\nonumber
\end{equation*}
valid for all $x \in supp \, f,$ implies the following
representation for the $q$-Fourier transform without usage of the
$q$-product:
\begin{equation}  \label{identity2}
F_q[f](\xi) = \int_{supp \,f} f(x) e_q^{ix \xi [f(x)]^{q-1}} dx.
\end{equation}
\par
The remaining of the paper is organised as follows: In Section
\ref{Preliminaries} we mention some roperties of $F_{q}$; In Section 3 we derive functional-differential
equations for the ${q}$-Fourier transform of $q$-Gaussians. Then,
based on the results of this Section, we show that the answer to the
above question is affirmative for all $q\geq 1$, and for two
particular values of $q<1$, namely for $q=1/2$ and $q=2/3.$ We also
show
that if $q<1,$ except two values mentioned above, $F_{q}$-transform of a $q$%
-Gaussian is no longer a $q^{^{\prime }}$-Gaussian, $\forall
q^{^{\prime }}<3. $ A relevant physical application of $F_q$ and the
functional-differential equations studied in Section 3 is addressed
in Section 4.

\section{Preliminaries}

\label{Preliminaries}

The following properties of $F_q$ follow immediately from its
representation (\ref{identity2}).

\begin{proposition}
\label{fqsimple} For any constants $a>0, \, b>0,$
\begin{enumerate}
\item $F_q[a f(x)](\xi)=a F_q [f(x)](\frac{\xi}{a^{1-q}});$
\item $F_q[f(bx)](\xi)=\frac{1}{b}F_q[f(x)](\frac{\xi}{b}).$
\end{enumerate}
\end{proposition}

Now we recall some facts related to $q$-Gaussians. Let $\beta $ be a
positive number. A function
\begin{equation}
G_{q}(\beta ;x)=\frac{\sqrt{\beta }}{C_{q}}e_{q}^{-\beta x^{2}},
\label{gaussian}
\end{equation}%
is called a $q$-Gaussian. The constant $C_{q}$ is the normalising
constant, namely $C_{q}=\int_{-\infty }^{\infty }e_{q}^{-x^{2}}dx,$
with explicit expression \cite{UmarovTsallisSteinberg}
\begin{equation}
C_{q}=\left\{
\begin{array}{ll}
{\ \frac{2}{\sqrt{1-q}}\int_{0}^{\pi /2}(cos\,t)^{\frac{3-q}{1-q}}dt}=\frac{2%
\sqrt{\pi }\,\Gamma \bigl({\frac{1}{{1-q}}}\bigr)}{(3-q)\sqrt{1-q}\,\Gamma %
\bigl({\frac{{3-q}}{{2(1-q)}}}\bigr)}, & -\infty <q<1, \\
\sqrt{\pi }, & q=1, \\
\frac{2}{\sqrt{q-1}}\int_{0}^{\infty }(1+y^{2})^{\frac{{-1}}{{q-1}}}dy=\frac{%
\sqrt{\pi }\,\Gamma \bigl(\frac{3-q}{2(q-1)}\bigr)}{\sqrt{q-1}\,\Gamma \bigl(%
{\frac{1}{{q-1}}}\bigr)}, & 1<q<3\,. \\
&
\end{array}%
\right.   \label{cq}
\end{equation}%
If $q<1$, then $G_{q}(\beta ;x)$ has a compact support $|x|\leq
K_{\beta }, $ where $K_{\beta }=(\beta (1-q))^{-1/2}.$ We use the
convention $K_{\beta }=\infty $ if $q\geq 1,$ since the support of a
$q$-Gaussian is not bounded in this case.

Note that $q$-exponentials possess the property $e_{q}^{z}\otimes
_{q}e_{q}^{w}=e_{q}^{z+w}$~\cite{Nivanen2003,qborges}. This
implies  the following proposition.

\begin{proposition}
\label{lem2} For all $q<3$ the $q$-Fourier transform of \, $e_q^{-\beta
x^2}, \, \beta >0,$ can be written in the form
\begin{equation}  \label{property1}
F_q[e_q^{-\beta |x|^2}](\xi) =\int_{-K_{\beta}}^{K_{\beta}} e_q^{-\beta
|x|^2+ix \xi}dx.
\end{equation}
\end{proposition}

\begin{corollary}
Let $q<3.$ Then
\begin{equation*}
F_q[e_q^{-\beta |x|^2}](\xi) = 2 \int_{0}^{K_{\beta}} e_q^{-\beta |x|^2}
cosh_q \left({\frac{x \xi}{[e_q^{-\beta |x|^2}]^{1-q}}}\right) dx, \,
\forall \, q,
\nonumber
\end{equation*}
where
\begin{equation*}
cosh_q(x)=\frac{e_q^x+e_q^{-x}}{2}.
\nonumber
\end{equation*}
\end{corollary}

The following assertion was proved in \cite{UmarovTsallisSteinberg}.

\begin{proposition}
\label{qgreater1} Let $1 \leq q <3.$ Then
\begin{equation}  \label{gausstransform21}
F_{q}[G_{q}(\beta; x)](\xi) = e_{q_{1}}^{- \beta_{\ast} \xi^2},
\end{equation}
where $q_{1}=\frac{1+q}{3-q}$ and $\beta_{\ast} = \frac{3-q}{8 \beta^{2-q}
C_{q}^{2(q-1)}}.$
\end{proposition}

\begin{proposition}
\label{qless1} Let $q<1.$ Then
\begin{equation*}
F_q[G_q(\beta,x)]= e_{q_1}^{-\beta_{\ast}|\xi|^2} (1-\frac{2}{C_q} Im
\int_0^{d_{\xi}} e_q^{b_{\xi}+i\tau}d\tau),
\end{equation*}
where $q_1 = (1+q)/(3-q),$ $C_q$ is the normalising constant and
$b_{\xi} +
i d_{\xi} = \frac{K_{\beta}\sqrt{\beta}- i \frac{\xi}{2 \sqrt{\beta}}}{%
[e_q^{-\frac{\xi^2}{4\beta}}]^{\frac{1-q}{2}}}. $
\end{proposition}

\textit{Proof.} The proof of this statement can be obtained applying
the Cauchy theorem, that is by integrating the function $e_q^{-\beta
z^2+i z \xi}$
over the closed contour $C=C_0 \cup C_1 \cup C_{-} \cup C_{+},$ where $%
C_p=(-K_{\beta} +pi,K_{\beta}+ip), \, p=0, 1,$ and $C_{\pm}=[\pm
K_{\beta}, \pm K_{\beta}+i].$ \eproof
\par
It follows from Propositions \ref{qgreater1} and \ref{qless1} that
\begin{equation*}
F_q[G_q(\beta,x)]=e_{q_1}^{-\beta_{\ast}|\xi|^2} + I_{(q<1)}(q) \, \,
T_q(\xi),
\nonumber
\end{equation*}
where $I_{(a,b)}(\cdot)$ is the indicator function of $(a,b)$, and
\begin{equation*}
T_q(\xi) = -\frac{2}{C_q} e_{q_1}^{-\beta_{\ast}|\xi|^2} Im \int_0^{d_{\xi}}
e_q^{b_{\xi}+i\tau}d\tau.
\nonumber
\end{equation*}
Thus for $q\geq 1$ $F_{q}$ transforms a $q$-Gaussian into a
$q_{1}$-Gaussian
with the factor $C_{q_{1}}\beta ^{-1/2}.$ However, for $q<1$, the tail $%
T_{q}(\xi )$ appears.

\begin{proposition}
\label{l1} For any real $q_1$, $\beta_1 >0$ and $\delta >0$ there exist
uniquely determined $q_2=q_2(q_1, \delta)$ and $\beta_2 = \beta_2(\delta,
\beta_1),$ such that
\begin{equation*}
(e_{q_1}^{-\beta_1 x^2})^{\delta} = e_{q_2}^{-\beta_2 x^2}.
\nonumber
\end{equation*}
Moreover, $q_2=\delta^{-1}(\delta - 1 +q_1),$ $\beta_2 = \delta \beta_1.$
\end{proposition}

\textit{Proof.} Let $q_{1} < 3,\beta _{1}>0$, and $\delta >0$ be any
fixed real numbers. For the equation,
\begin{equation*}
(1-(1-q_{1})\beta _{1}x^{2})^{\frac{\delta }{{1-q_{1}}}}=(1-(1-q_{2})\beta
_{2}x^{2})^{\frac{1}{{1-q_{2}}}}
\nonumber
\end{equation*}%
to be an identity, it is needed $(1-q_{1})\beta _{1}=(1-q_{2})\beta _{2},$ $%
1-q_{1}=\delta (1-q_{2}).$ These equations have a unique solution $%
q_{2}=\delta ^{-1}(\delta -1+q_{1}),$ $\beta _{2}=\delta \beta
_{1}.$ \eproof

\begin{corollary}
\label{cor20} $(e_{q}^{-\beta x^2})^q=e_{2-{\frac{1 }{q}}}^{-q\beta
x^2}.$
\end{corollary}

Now we introduce a sequence $q_n$ defined by the relation
\begin{equation}  \label{qk}
q_n = \frac{2q-n(q-1)}{2-n(q-1)}, \,
\end{equation}
where $-\infty < n < \frac{2}{q-1}-1$ if $1 < q <3$, and $n >
-\frac{2}{1-q}$ if $q \leq 1$. Notice that $q_n = 1$ for all $n=0,
\pm 1,...,$ if $q=1.$ Let $\mathbb{Z}$ be the set of all integer
numbers. Denote by $\mathbb{N}_q$ a subset of $\mathbb{Z}$ defined
as
\begin{equation*}
\mathbb{N}_{q} = \left\{
\begin{array}{ll}
\{n \in \mathbb{Z}: n < \frac{2}{q-1}-1\} , & \mbox{if $1 < q < 3,$} \\
\{n \in \mathbb{Z}: n > -\frac{2}{1-q}\}, & \mbox{if $q \leq 1$.}%
\end{array}
\right.
\nonumber
\end{equation*}

\begin{proposition}
\label{lem4} For all $n \in \mathbb{N}_q$ the relations

\begin{enumerate}
\item $(3-q_n)q_{n+1}=(3-q_{n-2})q_n, $

\item $2 C_{q_{n-2}}= \sqrt{q_n} \, (3-q_n) \, C_{q_n} $
\end{enumerate}

hold true.
\end{proposition}

\emph{Proof.} 1. It follows from the definition of $q_{n}$ that $%
q_{n+1}=(1+q_{n})/(3-q_{n}).$ This yields
\begin{equation}
(3-q_{n})q_{n+1}=1+q_{n}=(1+\frac{1}{q_{n}})q_{n}.  \label{step1}
\end{equation}%
Further, the duality relation $q_{k-1}+q_{k+1}^{-1}=2$ holds for all
$k\in \mathbb{N}_{q}.$ Applying it for $k=n-1$, we have
$1/q_{n}=2-q_{n-2}.$ Taking this into account in (\ref{step1}) we
arrive at statement 1).

2. For $q=1$, the relation 2) is reduced to the simple equality $2\sqrt{\pi }%
=2\sqrt{\pi }.$ Let $q\neq 1.$ Notice that if $1<q<3$ then,
$1<q_{n}<3$ for all $n\in \mathbb{N}_{q};$ if $q<1$ then, $q_{n}<1$
as well for all $n\in \mathbb{N}_{q}.$ Consider
$A_{n}=2C_{n-2}/C_{n}.$ Using the explicit form of $C_{q}$ given in
(\ref{cq}) and the duality relation $2-q_{n-2}=1/q_{n},$ in the case
$1<q<3$ one obtains
\begin{equation*}
A_{n}=\frac{\sqrt{q_{n}}\,\,\,\Gamma \left(\frac{1+q_{n}}{2(q_{n}-1)}\right)}{\frac{1}{%
2(q_{n}-1)}\,\,\,\Gamma \left(\frac{3-q_{n}}{2(q_{n}-1)}\right)}=\sqrt{q_{n}}(3-q_{n}).
\end{equation*}%
Further, if $q<1,$ then
\begin{equation*}
A_{n}=\frac{\sqrt{q_{n}}(3-q_{n})}{\frac{1+q_{n}}{2(1-q_{n})}}\,\frac{\Gamma
\left(\frac{3-q_{n}}{2(1-q_{n})}\right)}{\Gamma \left(\frac{1+q_{n}}{2(1-q_{n})}\right)}=\sqrt{%
q_{n}}(3-q_{n}),
\nonumber
\end{equation*}%
proving the statement 2). \eproof

\section{Main results}

\subsection{Functional differential equations}

Denote $g_q (\beta,\xi) = F_q [G_q (\beta, x)](\xi).$ For $\beta=1$, we use
the notation $g_q(\xi)=g_q(1,\xi).$ Let $Y(q,\xi)=F_q[e_q^{-x^2}](\xi).$ By
Proposition \ref{lem2},
\begin{equation*}
Y(q,\xi)=\int_{-K}^{K} e_q^{- |x|^2+ix \xi} dx,
\nonumber
\end{equation*}
where $K=K_1=\frac{1}{ \sqrt{1-q}}$ if $q<1,$ and $K=\infty,$ if $q \geq 1.$

\begin{lemma}
\label{lem5} For any $q<3$ and $\beta >0$ we have,

\begin{enumerate}
\item $g_q(\beta, \xi)= g_q(\frac{\xi}{(\sqrt{\beta})^{2-q}});$

\item $g_q(\xi)= \frac{1}{C_q}Y(q,C_q^{1-q}\xi).$
\end{enumerate}
\end{lemma}

\textit{Proof.} The proof follows from the properties of $F_q$ indicated in
Proposition \ref{fqsimple}.

These two formulae imply,
\begin{equation*}
F_q[G_q(\beta,x)](\xi)=\frac{1}{C_q}Y\left(q, \left(\frac{C_q}{\sqrt{\beta}}\right)^{1-q}
\frac{\xi}{\sqrt{\beta}}\right).
\nonumber
\end{equation*}
Moreover, $g_q(\beta,0)=1$, which implies $g_q(0)=1 \, \, \, \mbox{and} \,
\, \, Y(q,0)= C_q. $ Thus, it suffices to study $Y(q,\xi)$ in order to know
properties of the $q$-Fourier transform of $q$-Gaussians.

\begin{theorem}
\label{main1} Let $1 \le q < 3$ and $q_n, \, n \in \mathbb{N}_q,$
are defined in (\ref{qk}). Then $Y(q_n,\xi)$ satisfies the following
homogeneous functional-differential equation
\begin{equation}  \label{fde1}
2 \sqrt{q_n} \frac{\partial Y(q_n,\xi)}{\partial \xi} + \xi Y(q_{n-2}, \sqrt{%
q_n}\xi)=0;
\end{equation}
\end{theorem}

\emph{Proof.} Differentiating $Y(q,\xi )=\int_{-K}^{K}e_{q}^{-x^{2}+ix\xi }$
with respect to $\xi $, we have
\begin{equation*}
\frac{\partial Y(q,\xi )}{\partial \xi }=i\int_{-K}^{K}x(e_{q}^{-x^{2}+ix\xi
})^{q}dx.
\nonumber
\end{equation*}%
Further, integrating by parts, we obtain
\begin{equation}
\frac{\partial Y(q,\xi )}{\partial \xi }=\frac{-i}{2}%
\int_{-K}^{K}d(e_{q}^{-x^{2}+ix\xi })-\frac{\xi }{2}%
\int_{-K}^{K}(e_{q}^{-x^{2}+ix\xi })^{q}dx .  \label{parts}
\end{equation}%
It is not straightforward to see that the first integral vanishes if $q\geq 1.$
Applying Corollary \ref{cor20}, the second integral can be
represented in the form
\begin{equation}
\int_{-K}^{K}(e_{q}^{-x^{2}+ix\xi })^{q}dx =\frac{1}{\sqrt{q}}%
\int_{-K}^{K}e_{2-1/q}^{-x^{2}+ix\sqrt{q}\xi }dx =\frac{1}{\sqrt{q}}Y\left(2-%
\frac{1}{q},\,\sqrt{q}\xi \right).  \label{2ndintegral}
\end{equation}
Hence, for $q\geq 1$ the function $F_{q}[e_{q}^{-x^{2}}]$ satisfies the
functional-differential equation
\begin{equation}
2\sqrt{q}\frac{\partial Y(q,\xi )}{\partial \xi }+\xi Y(2-1/q,\sqrt{q}\xi
)=0.  \label{fdeq>1}
\end{equation}%
Now let $q=q_{n},n\in \mathbb{N}_{q}.$ Then taking into account the
relation $2-1/q_{n}=q_{n-2}$ we obtain (\ref{fde1}). \eproof

\begin{theorem}
\label{main2} Let $0< q < 1$ and $q \neq l/(l+1), l=1,2,....$ Then $%
Y(q_n,\xi)$ satisfies the following inhomogeneous functional-differential
equation
\begin{equation}  \label{fde2}
2 \sqrt{q_n} \frac{\partial Y(q_n,\xi)}{\partial \xi} + \xi Y(q_{n-2}, \sqrt{%
q_n}\xi)= r_{q_n} \xi^{\frac{1}{1-q_n}},
\end{equation}
where
\begin{equation}  \label{rq}
r_{q_n} = 2 \sqrt{q_n} \sin \frac{\pi}{2(1-q_n)} (1-q_n)^{\frac{1}{2(1-q_n)}%
}.
\end{equation}
\end{theorem}

\emph{Proof.} Assume that $q<1$ and $q\neq \frac{l}{l+1},l=1,2,....$ We
notice that if $q<1$ then the first integral on the right hand side of (\ref%
{parts}) does not vanish. Now it takes the form
\begin{equation*}
\int_{-K}^{K}d(e_{q}^{-x^{2}+ix\xi })=e_{q}^{-K^{2}+iK\xi
}-e_{q}^{-K^{2}-iK\xi }=2i\,\,Im\,\,e_{q}^{-K^{2}+iK\xi }.
\nonumber
\end{equation*}%
Since $supp\,e_{q}^{-x^{2}}=[-K,K],$ one has $e_{q}^{-K^{2}}=0.$
Hence,
\begin{equation*}
e_{q}^{-K^{2}+iK\xi }=0\otimes _{q}e_{q}^{iK\xi }=[i(1-q)K\xi ]^{\frac{1}{1-q%
}}.
\nonumber
\end{equation*}%
Further, taking into account $K=1/\sqrt{1-q},$ we obtain
\begin{equation*}
Im[i(1-q)K\xi ]^{\frac{1}{1-q}}=(1-q)^{\frac{1}{2(1-q)}}\sin \frac{\pi }{%
2(1-q)}\xi ^{\frac{1}{1-q}}.
\nonumber
\end{equation*}%
The expression in (\ref{2ndintegral}) for the second integral in the
right hand side of (\ref{parts}) is the same in the case of $q<1$. Hence, in this case $F_{q}[e_{q}^{-x^{2}}](\xi )$ satisfies
the functional-differential equation
\begin{equation}
2\sqrt{q}\frac{\partial Y(q,\xi )}{\partial \xi }+\xi Y(2-1/q,\sqrt{q}\xi
)=r_{q}\xi ^{\frac{1}{1-q}},  \label{fdeq<1}
\end{equation}%
where
\begin{equation*}
r_{q}=2\sqrt{q}(1-q)^{\frac{1}{2(1-q)}}\sin \frac{\pi }{2(1-q)}.
\nonumber
\end{equation*}%
Again, by taking $q=q_{n},n\in \mathbb{N}_{q},$ we arrive at the
functional-differential equation (\ref{fde2}).  \eproof

Now we consider the case $q=l/(l+1), \, l=1,2,...,$ excluded from Theorems \ref%
{main1} and \ref{main2}. In this case $K=\sqrt{l+1}$ and $Y(q,\xi )$ takes
the form
\begin{equation*}
Y(q,\xi )=F_{q}[e_{q}^{-x^{2}}](\xi )=\int_{-\sqrt{l+1}}^{\sqrt{l+1}}(1-%
\frac{1}{l+1}x^{2}+\frac{1}{l+1}ix\xi )^{l+1}dx.
\nonumber
\end{equation*}%
We use notation $P_{l+1}(\xi )=Y(\frac{l}{l+1},\xi )$ indicating the
dependence on $l$. Further, obviously
\begin{equation*}
2-{\frac{1}{q}}=\frac{l-1}{l},
\nonumber
\end{equation*}%
and consequently
\begin{equation*}
Y(2-1/q,\xi )=\int_{-\sqrt{l}}^{\sqrt{l}}(1-\frac{1}{l}x^{2}+\frac{1}{l}%
ix\xi )^{l}dx=P_{l}(\xi ).
\nonumber
\end{equation*}%
It is simple to see that $P_{l}(\xi )$ is a polynomial of even order, namely
of order $l$ if $l$ is even, and of order $l-1$ if $l$ is odd. Moreover, $%
P_{l}(\xi )$ is a symmetric function of $\xi $ and $P_{l}(0)=C_{\frac{l-1}{l}%
}>0.$ Let $\rho $ be a root of $P_{l}(\xi )$ closest to the origin.
We will consider $P_{l}(\xi )$ only on the interval $\xi \in \lbrack
-\rho ,\rho ],$ where it is positive.

\begin{theorem}
\label{main3} Let $q=\frac{2m-1}{2m}, \, m=1,2,....$ Then $Y(q,\xi)$
satisfies the functional-differential equation (\ref{fde1}).
\end{theorem}

\emph{Proof.} Assume $l+1=2m, m=1,2,....$ In this case $Y(q,\xi)=P_{2m}(\xi)$
is a polynomial of order $2m$ and $Y(2-1/q, \xi)=P_{2m-1}(\xi)$ is a
polynomial of order $2m-2$. Moreover, it is easy to check that in this case $%
r_q=0.$ Thus, $Y(q,\xi)$ satisfies the equation
\begin{equation}  \label{fdel=2m}
2 \sqrt{q} \frac{\partial Y(q,\xi)}{\partial \xi} + \xi Y(2-1/q,\sqrt{q}
\xi)(\xi)=0.
\end{equation}
It is easy to verify that this equation is consistent. \eproof

\begin{theorem}
\label{main4} Let $q=\frac{2m}{2m+1}, \, m=1,2,....$ Then $Y(q,\xi)$
satisfies neither the functional-differential equation (\ref{fde1}) nor (%
\ref{fde2}).
\end{theorem}

\emph{Proof.} Let $l=2m, m=1,2,....$ Then $Y(q,\xi)=P_{2m+1}(\xi)$ is a
polynomial of order $2m$, as well as $Y(2-1/q,\xi)=P_{2m}(\xi).$ Assume $%
Y(q,\xi)$ satisfies the equation (\ref{fde2}), which in this case
takes the form
\begin{equation}  \label{fdel=2m-1}
2 \sqrt{q} \frac{\partial Y(q,\xi)}{\partial \xi} + \xi P_{2m}(\xi)= \frac{%
(-1)^m}{(2m-1)^{m-{\frac{1 }{2}}}} \xi^{2m+1}.
\end{equation}
Clearly the derivative of a polynomial of order $2m$ can not be a
polynomial of order $2m+1.$ Analogously, $Y(q,\xi)$ cannot satisfy
equation (\ref{fde1}) either. \eproof

\subsection{Is the $q$-Fourier transform of a $q$-Gaussian a $q^{'}$-Gaussian?}

We shall now introduce the set of functions
\begin{equation}
\mathcal{G} =\bigcup_{q<3} \mathcal{G}_q, \, \, \, \mbox{where} \, \, \,%
\mathcal{G}_q = \{f: f(x)=a e_q^{-\beta x^2}, \, a>0, \, \beta>0\}.
\end{equation}

\begin{theorem}
\label{thmq>1} Let $1 \leq q_n < 3$. Then the following Cauchy problem for a
functional-differential equation
\begin{equation}  \label{eq100}
2 \sqrt{q_n} \frac{\partial Y(q_n,\xi)}{\partial \xi} + \xi Y(q_{n-2}, \sqrt{%
q_n}\xi)=0;
\end{equation}
\begin{equation}  \label{cauchy100}
Y(q_n,0)=C_{q_n},
\end{equation}
has a solution $Y(q_n,\xi) \in \mathcal{G}$ and this solution is specifically
\begin{equation}  \label{sol10}
Y(q_n,\xi)=C_{q_n} e_{q_{n+1}}^{-\frac{3-q_n}{8}\xi^2}.
\end{equation}
\end{theorem}

\emph{Proof.} It immediately follows from the representation that $%
Y(q_{n},0)=C_{q_{n}}.$ Furthermore,
\begin{equation*}
\frac{\partial Y(q_{n},\xi )}{\partial \xi }=-{\frac{1}{4}}%
(3-q_{n})\,\,C_{q_{n}}\xi \,\,\,\left( e_{q_{n+1}}^{-\frac{3-q_{n}}{8}\xi
^{2}}\right) ^{q_{n+1}},
\end{equation*}
\begin{equation}
Y(q_{n-2},\sqrt{q_{n}}\xi )=C_{q_{n-2}}e_{q_{n-1}}^{-q_{n}\frac{3-q_{n-2}}{8}%
\xi ^{2}}.  \label{eq101}
\end{equation}%
In addition, Corollary \ref{cor20} and relation 1) in Proposition
\ref{l1} imply that
\begin{equation}
\frac{\partial Y(q_{n},\xi )}{\partial \xi }=-{\frac{1}{4}}%
(3-q_{n})\,\,C_{q_{n}}\xi \,\,\,e_{q_{n-1}}^{-q_{n}\frac{3-q_{n-2}}{8}\xi
^{2}}.  \label{eq102}
\end{equation}%
Substituting (\ref{eq101}) and (\ref{eq102}) in equation
(\ref{eq100}), we obtain
\begin{equation}
(-\sqrt{q_{n}}C_{q_{n}}\frac{3-q_{n}}{2}+C_{q_{n-2}})e_{q_{n-1}}^{-\frac{%
q_{n}(3-q_{n})}{8}\xi ^{2}}=0.  \label{iden}
\end{equation}%
Now taking into account the second relation in Proposition \ref{l1} we
conclude that $Y(q_{n},\xi )$ in (\ref{sol10}) satisfies equation (\ref%
{eq100}). \eproof

\begin{corollary}
\label{cor}
Let $q_n \geq 1$. Then
\begin{equation}  \label{cor0}
F_{q_n}[G_{q_n}](\xi)= e_{q_{n+1}}^{-\frac{3-q_n}{%
8\beta^{2-q_n}C_{q_n}^{2(q_n-1)}}\xi^2}.
\nonumber
\end{equation}
\end{corollary}

\begin{remark}
Representation (\ref{cor0}) was obtained in
\cite{UmarovTsallisSteinberg} by the contour integration technique.
\end{remark}

\begin{theorem}
\label{thmq<1} Let $q_n < 1, \, n \in \mathbb{N}$ and $q_n \neq
m/(m+1), \, m=1,2....$ Then the Cauchy problem for a
functional-differential equation
\begin{equation}  \label{eq200}
2 \sqrt{q_n} \frac{\partial Y(q_n,\xi)}{\partial \xi} + \xi Y(q_{n-2}, \sqrt{%
q_n}\xi)=r_{q_n} \xi^{\frac{1}{1-q_n}},
\end{equation}
\begin{equation}  \label{cauchy200}
Y(q_n,0)=C_{q_n},
\end{equation}
has no solution in $\mathcal{G}$.
\end{theorem}

\emph{Proof.} Let $q_n<1, \, q_n \neq m/(m+1), m=1,2,....$ First, we notice
that a function with compact support can not solve equation (\ref{eq200}%
). It follows that a solution to (\ref{eq200}), $Y(q_n,\xi) \notin \mathcal{G%
}_q$ with $q<1,$ since any function in $\mathcal{G}_q$ for $q<1$ has compact
support. Now assume that there exists a $q= q (q_n)\ge 1$, such that $%
Y(q_n, \xi) \in \mathcal{G}_q$, that is
\begin{equation*}
Y(q_n,\xi)= A_{q_n} e_{q}^{-b(q_n)\xi^2},
\nonumber
\end{equation*}
where $A_{q_n}>0, \, b(q_n)>0$ are some real numbers. It follows from Eq.~(\ref%
{cauchy200}) that $A_{q_n}=C_{q_n}.$ Further, $Y(q_{n-2},\xi) \in \mathcal{G}%
_{q^{\ast}},$ where $q^{\ast}=2-1/q,$ that is $Y(q_{n-2},%
\xi)=C_{q_{n-2}}e_{q^{\ast}}^{-\beta(q_n)\xi^2}, \, \beta(q_n)>0.$ Then, for
$Y(q_n,\xi)$ to be consistent with Eq.~(\ref{eq200}), one has
\begin{equation*}
\frac{2}{1-q^{\ast}} + 1 = \frac{1}{1-q_n},
\nonumber
\end{equation*}
or $q^{\ast}=\frac{3q_n-2}{q_n}.$ Hence, $q=\frac{q_n}{2-q_n}<1,$ since $%
q_n<1.$ This contradicts the assumption that $q \ge 1.$
\eproof

Finally, considering the specific cases $q=\frac{1}{2}, \, \frac{2}{3},...,%
\frac{m}{m+1},....$ a direct computation shows that
\begin{equation*}
F_{\frac{1 }{2}} \left[e_{\frac{1 }{2}}^{-x^2}\right](\xi)= \frac{16\sqrt{2}}{15}\left(1-{%
\frac{5 }{16}} \xi^2 \right).
\nonumber
\end{equation*}
This function is non-negative for $|\xi| \le 4/\sqrt{5},$ so in this
interval we can associate it by $\frac{16 \sqrt{2}}{15}e_0^{-(5/16) \xi^2}
\in \mathcal{G}_0.$ The similar situation holds true in the case $q=2/3$ as
well yielding,
\begin{equation*}
F_{\frac{2 }{3}}\left[e_{\frac{2 }{3}}^{-x^2}\right](\xi) = \frac{32\sqrt{3}}{35}\left(1- {%
\frac{7 }{24}} \xi^2\right),
\nonumber
\end{equation*}
which is positive in the interval $(-\frac{2\sqrt{6}}{7}, \frac{2\sqrt{6}}{7}%
).$

Below we show that for all values of $q=3/4,4/5,...$ $F_q$-transform of $%
e_q^{-x^2}$ does not belong to $\mathcal{G}.$ First we obtain an explicit
form for $P_{m+1} (\xi) = F_{q}\left[e_q^{-x^2}\right]$. Recall that $P_{m+1}(\xi)$ is
a polynomial of order $m+1$ if $m+1$ is even. Otherwise it is a polynomial
of order $m.$

\begin{theorem}
Let $q=m/(m+1), m=1,2,....$ Then $Y(q,\xi)=P_{m+1}(\xi)$ is represented in
the form
\begin{equation}  \label{repr}
P_{m+1}(\xi) =\sum_{k=0}^{[\frac{m+1}{2}]} (-1)^k \left(
\begin{matrix}
m+1 \\
2k%
\end{matrix}
\right) (m+1)^{-k+{\frac{1 }{2}}} B\left(k+\frac{1}{2}, m-2k+2\right) \,
\xi^{2k},
\end{equation}
where $[x]$ means the integer part of $x$, and $B(a,b)$ is the Euler's
beta-function.
\end{theorem}

\emph{Proof.} Recall that if $q=\frac{m}{m+1}, m=1,2,...,$ then
$Y(q,\xi)$ has the form
\begin{equation*}
Y(q,\xi) = P_{m+1}(\xi) = \int_{-\sqrt{m+1}}^{\sqrt{m+1}} (1 - \frac{1}{m+1}%
x^2 + \frac{1}{m+1}i x \xi)^{m+1}dx.
\nonumber
\end{equation*}
We have
\begin{equation*}
P_{m+1}(\xi) = \sum_{k=0}^{m+1}\left(
\begin{matrix}
m+1 \\
k%
\end{matrix}
\right)D_k(m) \frac{(i\xi)^k}{(m+1)^k},
\nonumber
\end{equation*}
where
\begin{equation*}
D_k(m) = \int_{-\sqrt{m+1}}^{\sqrt{m+1}} (1 - \frac{1}{m+1}x^2)^{m-k+1}
x^{k} dx.
\nonumber
\end{equation*}
Explicitly $D_k(m)=0$ if $k$ is odd and $%
D_{2k}(m)=(m+1)^{k+1/2}B(k+1/2,m-2k+2)$ for
$k=0,...,[\frac{m+1}{2}]$ which leads to representation (\ref{repr}).
\eproof

\begin{theorem}
Let $q=m/(m+1), \, m=3,4,....$ Then $Y(q,\xi)\notin \mathcal{G}.$
\end{theorem}

\emph{Proof.} It follows from the representation (\ref{repr}) that the first
three terms of the polynomial $Y(q,\xi)$ are
\begin{eqnarray}
Y(q,\xi)&=P_{m+1}(\xi) \nonumber \\& = D_0(m) \left[1- (m+1)^2
\frac{B(\frac{3}{2}, m)}{B(\frac{1}{2}, m+2)} \xi^2 +
\frac{m(m+1)^3}{2} \frac{B(\frac{5}{2}, m-2)}{B(\frac{1}{2}, m+2)}
\xi^4 +...\right] \nonumber \\
\label{3terms1}
&=D_0(m) \left[1-\frac{2m+3}{8(m+1)} \xi^2 +
\frac{(2m+3)(2m+1)}{8(m+1)^2} \xi^4 + ...\right],
\end{eqnarray}
where
\begin{equation*}
D_0(m)=C_{\frac{m}{m+1}}=\sqrt{m+1}B(\frac{1}{2},m+2)= \frac{\sqrt{m+1}%
(m+1)!2^{m+2}}{(2m+3)!!} \, .
\nonumber
\end{equation*}

Now assume that $Y(q,\xi )\in \mathcal{G}_{q_{\ast }}$ for some $q_{\ast }<3.
$ Then $1/(1-q_{\ast })=(m+1)/2,$ or $q_{\ast }=(m-1)/(m+1).$
We have
\begin{equation*}
Y(q,\xi )=D_{0}(m)(1-\beta (m)\xi ^{2})^{[\frac{m+1}{2}]},
\nonumber
\end{equation*}%
where $\beta (m)>0$ and $|\xi |\leq 1/\sqrt{\beta (m)}.$ Applying the
binomial formula and keeping the first three terms, one has
\begin{equation}
Y(q,\xi )=D_{0}(m)\left[ 1-\frac{(m+1)\beta (m)}{2}\xi ^{2}+\frac{%
(m^{2}-1)[\beta (m)]^{2}}{8}\xi ^{4}+...\right] .  \label{3terms2}
\end{equation}%
Comparing the second and third terms of (\ref{3terms1}) and
(\ref{3terms2}), one obtains contradictory relations
\begin{equation*}
\beta (m)=\frac{2m+3}{4(m+1)^{2}}
\nonumber
\end{equation*}
and
\begin{equation*}
[\beta (m)]^{2}=\frac{%
(3m+3)(2m+1)}{(m-1)(m+1)^{3}}\neq \frac{(2m+3)^{2}}{16(m+1)^{4}}=[\beta
(m)]^{2},\,\,m=3,4,....
\nonumber
\end{equation*}%
which proves the statement. \eproof

\begin{remark}
The formula (\ref{repr}) for $q = 1/2$ and $q = 2/3$ gives
\begin{equation*}
F_{\frac{1 }{2}} \left[e_{\frac{1 }{2}}^{-x^2}\right](\xi)= \frac{16\sqrt{2}}{15}(1-{%
\frac{5 }{16}} \xi^2) = \frac{16 \sqrt{2}}{15}e_0^{-(5/16) \xi^2}, \, \xi
\in \left[-\frac{4\sqrt{5}}{5},\frac{4\sqrt{5}}{5}\right],
\nonumber
\end{equation*}
and
\begin{equation*}
F_{\frac{2 }{3}}\left[e_{\frac{2 }{3}}^{-x^2}\right](\xi) = \frac{32\sqrt{3}}{35}(1- {%
\frac{7 }{24}} \xi^2) = \frac{32\sqrt{3}}{35} e_0^{-{\frac{7 }{24}} \xi^2},
\, \xi \in \left[-\frac{2\sqrt{6}}{7}, \frac{2\sqrt{6}}{7}\right].
\nonumber
\end{equation*}
Both functions belong to $\mathcal{G}_0.$
\end{remark}

\begin{remark}
If $q=1$ then the Cauchy problem (\ref{fde1}), (\ref{cauchy100})
reads
\begin{equation*}
2Y^{^{\prime }}(\xi )+\xi Y(\xi )=0,\,\,\,Y(0)=\sqrt{\pi },
\nonumber
\end{equation*}%
and its unique solution is $Y(\xi )=\sqrt{\pi }e^{-\xi ^{2}/4}.$
Besides from Corollary \ref{cor} we obtain
\begin{equation*}
F\left[\frac{\sqrt{\beta }}{\sqrt{\pi }}e^{-\beta x^{2}}\right]=e^{-\frac{1}{4\beta }%
\xi ^{2}}.
\nonumber
\end{equation*}%
The density of the standard normal distribution corresponds to $\beta =1/2,$
giving the characteristic function of the classic Gaussian.
\end{remark}

\section{Some applications to the porous medium equation}
\label{applications}

In this Section we discuss some applications of the $q$-Fourier
transform $F_{q}$ to nonlinear models of partial differential equations.
First we verify that the theorems proved in Section 3 imply that
$F_{q}$ transfers a $q$-Gaussian into a $q_{1}$-Gaussian if $q\geq
1,~~q_{1}=(1+q)/(3-q).$ Moreover, as shown in
\cite{UmarovTsallis2008}, the operator $F_{q}:G_{q}\rightarrow
G_{q_{1}}$ for $q>1$ is invertible. These two facts have been
essentially used in \cite{UmarovTsallisSteinberg,UmarovTsallis2007,Vignat}
for the proof of $q$-versions of the central limit theorem. Another
application of $F_{q}$, as sketched below, shows that it can be used
for establishing a relation between the porous medium equation
and a nonlinear ordinary differential equation (ODE) similar to the
usual Fourier transform.

The classic Fourier transform reduces the Cauchy problem for linear
partial differential equations of the form
$u_t(t,x)=A(D_{x})u(t,x)~~t>0,~~x\in
R^{n},
~~ u(0,x)=\varphi (x),
$ where $D_{x}=(D_{1},...,D_{n}),~~D_{j}=-i\frac{\partial }{\partial x_{j}}%
,j=1,...,n,$ and $A(D_{x})$ is an elliptic differential operator, to
an associated linear ODE with parameter $\xi \in R^n$. In the
particular case of $n=1$ and $A(D_{x})=\frac{d^{2}}{dx^{2}}$ for the
Fourier image $\hat{u}(t,\xi )$ of a solution $u(t,x)$, we have a
dual differential equation
\begin{equation}
\hat{u}_{t}^{^{\prime }}(t,\xi )=-\xi ^{2}\hat{u}(t,x),~~  \label{dualdifeq}
\hat{u}(0,\xi )=\hat{\varphi}(\xi ),
\end{equation}%
where $\xi \in R^{1}$ is a parameter. This case corresponds to the
Fokker-Planck equation for a Brownian motion without drift
\cite{Risken1989}.

We now demonstrate the similar role of $F_{q}$ in a simple model case, corresponding to
the cellebrated \textit{porous medium equation} in the
superdiffusion regime ubiquitously found in physical phenomena
\cite{CarrilloToscani1999,Otto2001,Muskat1937,TsallisBukman1996,Vazquez}
(and references therein) \footnote{The monograph \cite{Vazquez}
contains different approaches to the solution of the porous medium
equation.}. Consider the following non-linear diffusion equation
with a singular diffusion coefficient,
\begin{equation}
\frac{\partial U}{\partial
t}=(U^{1-q}U_{x})_{x},~~t>0,~~x\in R^{1},~~q>1. \label{nonlinde}
\end{equation}%
We look for a solution in the similarity set
$G_q^{\ast}=\{U(t,x):U(t,x)= t^{a}G_{q}(\beta ;t^{b}x), \, a=a(q),
\, b=b(q) \in R^1, \, \beta=\beta(q) >0 \},$ where $a$ and $\beta$
do not depend on $t$ and $x.$

\begin{proposition}
\label{carpde} Suppose $U(t,x)\in G_q^{\ast}$ is a solution to
Eq.~(\ref{nonlinde}).
 Then its $q$-Fourier transform
$\hat{U}_{q}(t,\xi )=F_q[U(t,x)](\xi)$ satisfies the following
nonlinear ordinary differential equation with parameter $\xi$
\begin{equation}
(\hat{U}_{q})_{t}^{^{\prime }}=-\frac{B(\beta ,q)\xi ^{2}}{t^{\frac{q-1}{3-q}%
}}(\hat{U}_{q})^{q_{1}},~~t>0,
\end{equation}
where $B(\beta,q)=\frac{2-q}{4\beta^{2-q}C_q^{q-1}}$ and
$q_1=\frac{1+q}{3-q}.$
\end{proposition}

\emph{Proof.} Let $U \in G_q^{\ast}$ be a solution to
(\ref{nonlinde}), i.e. for some $a=a(q)$ and $\beta=\beta(q)$ it has
representation $U(t,x) = t^{a}G_{q}(\beta ;t^{a}x)$. Then, it
follows from Proposition 2.1 that,
\begin{eqnarray}
\hat{U}_{q}(t,\xi )&=F_{q}[U(t,x)](\xi ) \nonumber \\&=F_{q}[G_{q}(\beta ;x)]\left(\frac{\xi }{%
t^{a(2-q)}}\right)=\frac{1}{C_{q}}Y\left(q, \left( \frac{\sqrt{\beta
}}{C_{q}}\right) ^{q-1}\frac{\xi }{\sqrt{\beta }t^{a(2-q)}}\right),
\nonumber
\end{eqnarray}%
where $Y(q,\xi )$ is a solution to equation (\ref{eq100}).
Computing the derivative of $\hat{U}_q(t,x)$ in variable $t,$
taking into account that $a=-1/(3-q)$ (see, e.g. \cite{Vazquez}),
and using equation (\ref{eq100}), we obtain
\[
(\hat{U}_q)_t = -\frac{2-q}{4 \beta^{2-q}C_q^{2(q-1)}} \xi^2
(\hat{U}_q)^{q_1},
\]
where $q_1=(1+q)/(3-q).$ \eproof

The inverse statement, given in the following formulation, is also
true.
\begin{proposition}
\label{relation2} Suppose $V(t,\xi), \, V(0,\xi)=1,$ is a solution
to ODE with parameter $\xi$
\begin{equation}
\label{ODE}
V^{^{\prime }}=-\frac{B(\beta ,q)\xi ^{2}}{t^{\frac{q-1}{3-q}%
}}V^{q_{1}},~~t>0,
\end{equation}
where $B(q,\beta)$ and $q_1$ are
as in Proposition \ref{carpde}. Then its inverse $q$-Fourier
transform $U(t,x)=F_q^{-1}[V(t,\xi)](x)$ exists and satisfies
equation (\ref{nonlinde}).
\end{proposition}

\emph{Proof.} By separation of variables of (\ref{ODE}) one can
verify that its solution
\[
V(t,\xi)= e_{q_{_1}}^{-\frac{3-q}{8\beta^{2-q}C_q^{q-1}}\Big(\xi t^{\frac{2-q}{3-q}}\Big)}.
\]
By Theorem 0.6 of paper \cite{UmarovTsallis2008} the inverse
$q$-Fourier transform for $V(t,\xi)$ exists, and by virtue of
Propositions \ref{fqsimple} and \ref{qgreater1} it has the
representation
\begin{equation}
U(t,x)=\frac{1}{t^{\frac{1}{3-q}}}G_{q}\left(\beta (q);\frac{x}{t^{\frac{1}{3-q}}}%
\right),\,\,\mbox{where}\,\,\beta (q)=\frac{1}{\left[2(3-q)C_{q}^{\frac{1}{q-1}}\right]^{%
\frac{2}{3-q}}}.  \label{solutionpme}
\end{equation}%
The latter is a solution to (\ref{nonlinde}); see \cite{Vazquez}.
\eproof

Notice that, if the initial condition is given in the form
$U(0,x)=\delta (x)$ with the Dirac's delta, and $q=1,$ then we
obtain equation (\ref{dualdifeq}) ($\hat{\varphi}(\xi)\equiv 1$), in
which $\beta =1/4,~~B(\beta ,1)=4\beta =1.$

In order to study price fluctuations in stock markets it was introduced
in~\cite{Borland} a stochastic process $X_t$ defined from a
stochastic differential equation $dX_t=\tau X_t + \sigma d
\Omega_t,$ where $\tau$ and $\sigma$ are the drift and volatility
coefficients respectively, and $\Omega_t$ is a solution to the Ito
stochastic differential equation
\begin{equation}
d\Omega_t = [P(\Omega_t)]^{\frac{1-q}{2}}dB_t \, .
\end{equation}
In this equation $B_t$ is a
Brownian motion, and $P$ is a $q$-Gaussian distribution function.
The corresponding Fokker-Planck type equation reads
\[
\frac{\partial V(x,t|x',t')}{\partial
t}=([V(x,t|x',t')]^{2-q})_{xx},
\]
which can easily be reduced to the form (\ref{nonlinde}). From the
financial applications point of view it is important to know the
properties of the stochastic process $X_t,$ since it can be
considered as a $q$-alternative to the Brownian motion. One can
effortlessly verify that if $U(t,x)$ is a solution to equation
(\ref{nonlinde}) for $t>0$ with an initial condition $U(0,x)=f(x),$
then a solution $V(t,x), ~ t>t'$ to the same equation
(\ref{nonlinde}) considered for $t>t'$ with an initial condition
$V(t',x)=f(x)$ can be represented in the form $V(t,x)=U(t-t',x), ~
t>t'.$ It follows that $X_t$ has stationary increments.

Concluding the discussion we note that equation (\ref{solutionpme})
corresponds to the solution obtained from an ansatz
\cite{TsallisBukman1996} which has been at the base of the
generalised Central Limit Theorem presented in
\cite{UmarovTsallisSteinberg}.

\section{Conclusion}

Summarising, we have the following general picture for the
$q$-Fourier transform of $q$-Gaussians.
\begin{itemize}
\item[1.]
The case $1 \le q < 3:$ for these values of $q$ \subitem  (1a) the
$q$-Fourier transform acts as $F_q: \mathcal{G}_q \to
\mathcal{G}_{q^{'}};$ \subitem (1b) the relation between $q$ and
$q^{'}$ is given by $q^{'}=\frac{1+q}{3-q}.$
\item[2.] The cases $q={1 \over 2}$ or $q={2 \over 3}:$ ~ for these two values of $q$
the operator acts as $F_q : \mathcal{G}_q \to \mathcal{G}_0,$
however the relationship (1b) is failed.
\item[3.] The case $q < 1,$ but $ q \neq {1 \over 2}, {2 \over 3}:$
~ in this case (1a) is failed as well in the sense that there is no
$q^{'}$ such that the $q$-Fourier transform of a $q$-Gaussian would
be a $q^{'}$-Gaussian.
\end{itemize}
The lesson we have learnt from the above analysis is that the
operator $F_q$ defined by formula (\ref{Fourier}) (or, the same, by
formula (\ref{identity2})) is rich in content and applicable only if
$q \in [1,3).$ Its important application is given in
\cite{UmarovTsallisSteinberg}  in the prove of the $q$-central limit
theorem and in \cite{UmarovTsallisGell-MannSteinberg} in conjunction
with $(q,\alpha)$-stable distributions. Another application of $F_q$
to the porous medium equation and related stochastic differential
models with time dependent variance are discussed in
Section~\ref{applications} of the current paper. What concerns the
case $q<1$, the $q$-Fourier transform defined by formula
(\ref{Fourier}) is not meaningful. An appropriate alternative
definition of $F_q$ in this case is remaining a challenging open
question.

\bigskip
We acknowledge C Tsallis for several comments on the subjects mentioned in
this article.

\end{document}